# High-order spectral singularity

H. S. Xu, L. C. Xie, and L. Jin[*]
*School of Physics, Nankai University, Tianjin 300071, China*

Exceptional point and spectral singularity are two types of singularity that are unique to non-Hermitian systems. Here, we report the high-order spectral singularity as a high-order pole of the scattering matrix for a non-Hermitian scattering system, and the high-order spectral singularity is a unification of the exceptional point and spectral singularity. At the high-order spectral singularity, the scattering coefficients have high-order divergence and the scattering system stimulates high-order lasing. The wave emission intensity is polynomially enhanced, and the order of the growth in the polynomial intensity linearly scales with the order of the spectral singularity. Furthermore, the coherent input controls and alters the order of the spectral singularity. Our findings provide profound insights into the fundamentals and applications of high-order spectral singularities.

## I. INTRODUCTION

Exceptional point (EP) in discrete spectra and spectral singularity (SS) in continuous spectra are concepts unique to non-Hermitian systems [1–22]. The SS is a pole of the scattering matrix where the scattering coefficients diverge at the real frequency. Consequently, at the SS, an optical system acts as a laser and its time-reversed counterpart acts as a coherent perfect absorber (CPA) [23–28]. Furthermore, incident-direction-dependent wave emissions are possible [29,30]. By contrast, the EP is a singularity where eigenvalues and their associated eigenstates coalesce. Consequently, a non-Hermitian system at the EP is non-diagonalizable and can be reduced only into a Jordan block form [31–36]. The intrinsic properties of the EP, in particular its geometric topology and unidirectionality, stimulate many useful applications including high-precision sensing [37–48], topological energy transfer [49,50], directional lasing [51], and asymmetric wave propagation [52–54] in photonics and beyond.

The coalescence of two solutions of a wave operator with a purely incoming boundary condition, named the CPA EP, has recently been proposed [55] and experimentally demonstrated [56–59]. The CPA EP has an anomalously broadened quartic absorption line shape. Notably, the time-reversed CPA EP is a second-order SS. In general, the $m$th-order SS is an $m$th-order pole of the scattering matrix at the real frequency. The wave vector $k = k_s$ is an $m$th-order SS when any scattering coefficient $s_{qp}(k)$ experiences an $m$th-order divergence $s_{qp}^{-1}(k_s) = 0$ and

$$\left.\frac{d^l}{dk^l} s_{qp}^{-1}(k)\right|_{k=k_s} \begin{cases} = 0 & (l < m) \\ \neq 0 & (l = m) \end{cases}, \quad (1)$$

where $s_{qp}(k)$, from the $q$th row and the $p$th column of the scattering matrix $S(k)$, is the output in port $q$ for the input in port $p$. The reflections are the diagonal terms of $S(k)$ and the transmissions are the off-diagonal terms of $S(k)$. However, the mechanism through which a high-order SS forms is an open question, and the associated time-evolution dynamics of high-order lasing are unclear. These unknowns are hindering development of the fundamentals and applications of high-order SSs in physics. In this work, we find that the scattering center at the $m$th-order EP coalesces the ports at the first-order SS to create the $m$th-order SS. This results in a high-order laser that provides the high-order lasing with input-dependent polynomial enhancement of the emission intensity.

The remainder of the paper is organized as follows. In Sec. II, the scattering formalism for a general scattering center is introduced and the scattering matrix is provided. In Sec. III, the formation mechanism of the high-order SS is elucidated. The scattering center at the high-order EP coalesces the ports at the first-order SS and generates the high-order SS. In Sec. IV, the lasing dynamics for the high-order SS are discussed. In Sec. V, we show the effect of EP. Notably, the coherent input can alter the order of lasing. In Sec. VI, the high-order SS in the coupled fiber taper system and the high-order CPA as a time-reversed laser are discussed. The conclusions are summarized in Sec. VII.

## II. SCATTERING FORMALISM

Figure 1 illustrates a general scattering system, which includes $N$ coupled resonators as the scattering center. Each individual resonator is connected to a resonator array as the port. The Hamiltonian of the scattering center is $H_c$. The ports have the uniform coupling $-\kappa$, the resonant frequency of the clockwise or counterclockwise mode is $\omega_c$, and the resonators' quality factor is $Q$. To investigate the system's scattering properties, we consider the lossless ports where the coupling strength is far larger than the intrinsic resonator dissipation $\kappa \gg \omega_c/Q$.

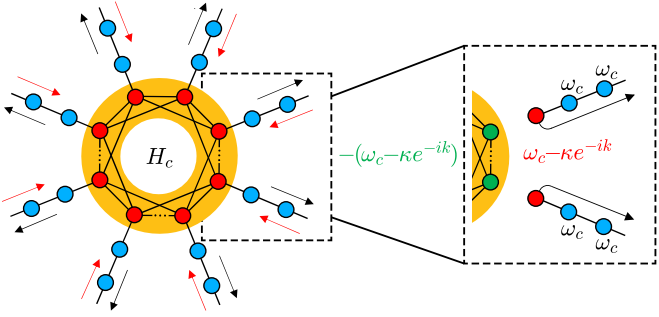

FIG. 1. Schematic of a discrete system. The port sites are shown in cyan. The scattering center sites are shown in red. Inset: The scattering system is formally decomposed into the ports ending with gain $\omega_c - \kappa e^{-ik}$, which are coupled through the connection Hamiltonian $h_c(k)$ as shaded in orange.

In temporal coupled mode theory [60–62], the time-evolution dynamics are governed by the equations of motion. For the resonator $j \neq 0$ of port $q$, we have

$$i\frac{d}{dt}\psi_q(j,t) = \omega_c \psi_q(j,t) - \kappa \psi_q(j-1,t) - \kappa \psi_q(j+1,t), \quad (2)$$

where $\psi_q(j,t)$ is the mode amplitude for resonator $j = 0, 1, 2, \ldots$ of port $q$. For the scattering center, we have

$$i\frac{d}{dt}\begin{pmatrix}\psi_1(0,t)\\ \vdots \\ \psi_q(0,t) \\ \vdots \\ \psi_N(0,t)\end{pmatrix} = H_c \begin{pmatrix}\psi_1(0,t)\\ \vdots \\ \psi_q(0,t) \\ \vdots \\ \psi_N(0,t)\end{pmatrix} - \kappa \begin{pmatrix}\psi_1(1,t)\\ \vdots \\ \psi_q(1,t) \\ \vdots \\ \psi_N(1,t)\end{pmatrix}. \quad (3)$$

Notably, the mode amplitude $\psi_q(0,t)$ for the resonator $j = 0$ of port $q$ is the mode amplitude for the resonator $q$ of the scattering center from the configuration in Fig. 1.

In the elastic scattering process, $\psi_q(j,t) = f_q(j)e^{-i\omega t}$ is the mode amplitude for resonator $j$ of port $p$. For the input in port $p$, the stationary mode amplitude $f_q(j) = \delta_{qp}e^{-ikj} + s_{qp}(k)e^{ikj}$ is a superposition of the incoming ($e^{-ikj}$) and outgoing ($e^{ikj}$) plane waves, where $k \in [0, \pi]$ is the wave vector for a monochromatic incident wave and $\delta_{qp}$ is the Kronecker delta. We also have $\psi_q(0,t) = [\delta_{qp} + s_{qp}(k)]e^{-i\omega t}$ and $\psi_q(1,t) = [\delta_{qp}e^{-ik} + s_{qp}(k)e^{ik}]e^{-i\omega t}$. After substituting the mode amplitude shown in Eq. (2), we obtain the dispersion relation of the resonator array $\omega = \omega_c - 2\kappa \cos k$. After substituting the mode amplitudes and dispersion relation into Eq. (3), we obtain

$$\begin{pmatrix}s_{1p}(k)\\ \vdots \\ s_{qp}(k) \\ \vdots \\ s_{Np}(k)\end{pmatrix} = -\frac{H_c - (\omega_c - \kappa e^{ik})I_N}{H_c - (\omega_c - \kappa e^{-ik})I_N}\begin{pmatrix}\delta_{1p}\\ \vdots \\ \delta_{qp} \\ \vdots \\ \delta_{Np}\end{pmatrix}. \quad (4)$$

Notably, $[s_{1p}(k), \ldots, s_{Np}(k)]^T$ is the $p$th column of scattering matrix $S(k)$, whereas $(\delta_{1p}, \ldots, \delta_{Np})^T$ is the $p$th column of the $N \times N$ identity matrix $I_N$. Therefore, we obtain the scattering matrix as

$$S(k) = -\frac{H_c - (\omega_c - \kappa e^{ik})I_N}{H_c - (\omega_c - \kappa e^{-ik})I_N}, \quad (5)$$

where $k \in [0, \pi]$ is the wave vector for monochromatic incidence and $I_N$ is the $N \times N$ identity matrix [63,64].

## III. HIGH-ORDER SS

The scattering system can be formally decomposed into ports ending with the gain $\omega_c - \kappa e^{-ik}$ and the connection Hamiltonian $h_c(k) = H_c - (\omega_c - \kappa e^{-ik})I_N$ (inset of Fig. 1). In this sense, each port is a one-port scattering system at the first-order SS, and all the ports are coupled by the connection Hamiltonian. The scattering coefficient divergence requires that $h_c(k)$, the denominator of $S(k)$, has an eigenvalue $\epsilon_k = 0$ at the critical $k = k_s$. However, in what type of system does the high-order divergence occur? We find that $h_c(k)$ (or equivalently $H_c$) at the EP generates the high-order divergence in $S(k)$.

Notably, $h_c(k)$ under a (nonsingular) similar transformation $E_N(k) = \Theta^{-1}h_c(k)\Theta$ is either a *diagonal* canonical form for a diagonalizable $H_c$ or a *Jordan* canonical form for a nondiagonalizable $H_c$. The operator $\Theta$ is $k$ independent because $H_c$ is $k$ independent. For a diagonalizable $H_c$, $\epsilon_k$ presents in the *diagonal* canonical form $E_N(k)$ of $h_c(k)$; in this case, the first-order divergence $\epsilon_k^{-1}$ appears in $h_c^{-1}(k)$, and consequently, the first-order divergence appears in $S(k)$. The wave vector $k$ is then the first-order SS [8]. For a nondiagonalizable $H_c$, the $m$th-order Jordan block $J_m$ presents in the Jordan canonical form $E_N(k)$ of $h_c(k)$ if $m$ eigenvalues of $h_c(k)$ coalesce to $\epsilon_k = 0$; in this case, the divergences from the first-order $\epsilon_k^{-1}$ to the $m$th-order $\epsilon_k^{-m}$ appear in $h_c^{-1}(k)$, and consequently, the $m$th-order divergence appears in $S(k)$. This is the $m$th-order SS. More specifically, the $m$th-order Jordan block has the form of

$$J_m = \begin{pmatrix}\epsilon_k & & & \\ 1 & \ddots & & \\ & \ddots & \ddots & \\ & & 1 & \epsilon_k\end{pmatrix}, \quad (6)$$

where the eigenvalue $\epsilon_k$ is the diagonal term. The unfilled elements are zeros. Without loss of generality, we assume that $J_m$ appears in the upper left corner of $E_N(k)$. The inverse of the $m$th-order Jordan block $J_m^{-1}$ also appears in the upper left corner of $E_N^{-1}(k)$ and $J_m^{-1}$ is a triangular matrix in the form of

$$-\begin{pmatrix}(-\epsilon_k)^{-1} & & & & \\ (-\epsilon_k)^{-2} & (-\epsilon_k)^{-1} & & & \\ \vdots & \vdots & \ddots & & \\ (-\epsilon_k)^{-m+1} & (-\epsilon_k)^{-m+2} & \cdots & (-\epsilon_k)^{-1} & \\ (-\epsilon_k)^{-m} & (-\epsilon_k)^{-m+1} & \cdots & (-\epsilon_k)^{-2} & (-\epsilon_k)^{-1}\end{pmatrix}. \quad (7)$$

The divergences from the first-order $\epsilon_k^{-1}$ to the $m$th-order $\epsilon_k^{-m}$ appear in $J_m^{-1}$.

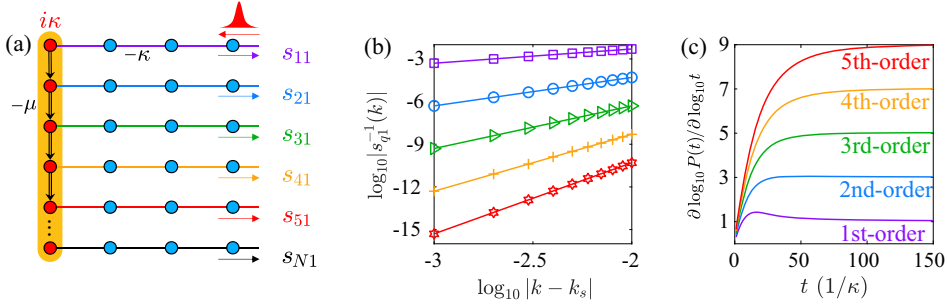

FIG. 2. (a) $N$-port scattering system at the $N$th-order SS. (b) Scaling behaviors near $k_s = \pi/2$ and (c) intensity growth in the outputs of the ports $q \in [1, 5]$ for the input in the first port $p = 1$ in the situation where the unidirectional coupling equals the uniform coupling $\mu = \kappa = 1$.

When the energy levels coalesce to the zero-energy $\epsilon_{k_s} = 0$, the inverse of the $m$th-order Jordan block $J_m^{-1}$ includes the divergences ranging from the first-order $\epsilon_{k_s}^{-1}$ to the $m$th-order $\epsilon_{k_s}^{-m}$. The $m$th-order divergence is the highest order of divergence that is associated with the inverse of the $m$th-order Jordan block $J_m^{-1}$. Consequently, the $m$th-order divergence $\epsilon_k^{-m}$ may appear in the scattering coefficients, and this is the $m$th-order SS.

The interplay between the gain $\omega_c - \kappa e^{-ik}$ and the uniform coupling $-\kappa$ supports the steady-state outgoing plane wave solution in the port. The scattering center $H_c$ at the $m$th-order EP with the complex frequency $\omega_c - \kappa e^{-ik}$ coalesces the ports at identical first-order SS, which differs from the coincidence of different first-order SSs [55]. The coalescence maintains the outgoing plane wave solution and creates the $m$th-order SS. In projection theory, an effective non-Hermitian Hamiltonian equivalently describes the original system at the steady state in a reduced dimension [65]. The steady-state solution at an SS of any order is the outgoing plane wave [66,67]; thus, a scattering system for which every individual port has been equivalently substituted by self-energy $-\kappa e^{ik}$ on the scattering center site yields an effective non-Hermitian Hamiltonian $H_c - \kappa e^{ik} I_N = h_c(k) + \omega I_N$, where $\omega = \omega_c - 2\kappa \cos k$ is the frequency of the monochromatic incidence. Consequently, the $m$th-order SS is also the $m$th-order EP of the whole scattering system, and the scattering matrix $S(k)$ is at the $m$th-order EP.

Non-Hermitian systems with coalesced zero energy at the third- [40], fourth- [68], fifth- [69,70], and sixth-order EPs [43] as the connection Hamiltonian $h_c(k)$ create the third-, fourth-, fifth-, and sixth-order SSs, respectively. If $h_c(k)$ is a non-Hermitian system with coalesced zero energy at the $N$th-order EP [71], the $N$th-order SS forms at the resonant frequency. Notably, any system at the $N$th-order EP relates to Jordan block $J_N$ under a (nonsingular) similar transformation. Therefore, Jordan block $J_N$ is the core of the $N$th-order EP. Consequently, the properties of the scattering system at the $N$th-order SS are acquired from the prototypical configuration presented in Fig. 2(a); the connection Hamiltonian $H_c$ is selected $-J_N$, which only consists of unidirectional couplings $-\mu$ (arrows) [72]. The unidirectional couplings are realizable through the resonator coupled to two point scatterers [51], the S-bend (Taiji) resonator [73–75], and the partially reflecting end mirror in the waveguide [57]. To create the high-order SS, the unidirectional couplings do not necessarily need to be uniform because the Jordan block form of the Hamiltonian is on the exceptional surface [71]. And the high-order SSs created from the unidirectional couplings are robust to disorder [76].

The derivatives of the scattering coefficients identify the order of the SS. The reflection coefficient is

$$s_{pp}(k) = -\frac{i + e^{ik}}{i + e^{-ik}}. \tag{8}$$

The transmission coefficients are $s_{qp}(k) = 0$ for the input in the port $q < p$ and

$$s_{qp}(k) = -\left(\frac{\mu}{\kappa}\right)^{q-p} \frac{2i \sin k}{(i + e^{-ik})^{q-p+1}}, \tag{9}$$

for the input in the port $q > p$. Then, the scattering matrix $S(k)$ is at the $N$th-order EP. For the input in port 1, the $q$th-order SS appears in port $q$ because $s_{q1}(k)$ have $q$th-order divergence at the resonant momentum $k_s = \pi/2$ and the scaling law near $k = k_s$ shown in Fig. 2(b) is

$$\left|s_{q1}^{-1}(k)\right| \approx \left|\frac{1}{q!} d^q s_{q1}^{-1}(k)/dk^q\Big|_{k=k_s} (k - k_s)^q\right| \propto |k - k_s|^q. \tag{10}$$

## IV. DYNAMICS AT THE HIGH-ORDER SS

The steady-state solution for the SS of any order is the purely outgoing waves, which cannot distinguish the order of SS; however, the time evolution of a Gaussian wave packet uncovers the order of the high-order lasers and reveals the enhanced wave emission.

In general, the propagating Gaussian wave packet in the wave vector $k$ space can be written as

$$\varphi(k, t) = \frac{\sqrt{\sigma}}{\sqrt[4]{\pi}} e^{-\sigma^2(k-k_c)^2/2} e^{-i(k-k_c)N_t}, \tag{11}$$

where $k_c$ is the central wave vector, $2\sqrt{\ln 2}/\sigma$ is the full width at half maximum, and $N_t$ indicates the center position of the Gaussian wave packet in the propagation. We focus on the resonant input at $k_c = k_s = \pi/2$. Thus, the deformation in the propagation of the wave packet is ignored. In this case, the wave propagating velocity $v = d\omega/dk|_{k=k_c}$ is $2\kappa$ and $N_t \propto t$. The wave packet localizes in the region $|k - k_c| < 2\sqrt{2 \ln 2}/\sigma$. The wave vector $k_c$ component dominates in the wave packet at large $\sigma$. Thus, the wave packet dynamics simulate the dynamic behavior of the plane wave at the wave vector $k_c$.

The output wave in port $q$ can be obtained using the scattering matrix. Thus, the scattered wave packet $\varphi_{qp}$ in the

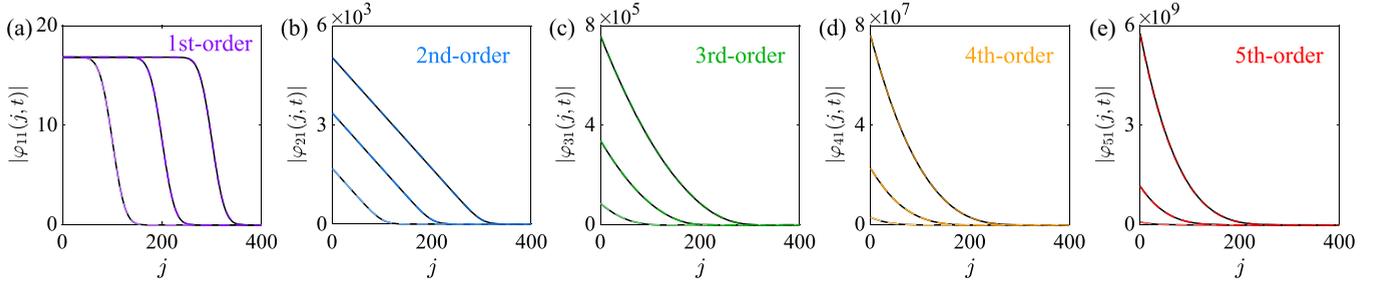

FIG. 3. The profile of time evolution for the injected Gaussian wave packet. (a)–(e) correspond to the $|\varphi_{q1}(j,t)|$ with $q \in [1, 5]$. The colored lines from lightness to darkness represent the simulations at $t = 50, 100, 150$ and the black line represents analytical results. In the simulation, the system has five ports and each port has 401 sites, and the site number $j$ ranges from 0 to 400. The Gaussian wave packets have $\sigma = 20$ and $k_c = \pi/2$.

momentum space is written as

$$\varphi_{qp}(k,t) = s_{qp}(k)\varphi(k,t). \quad (12)$$

For $s_{q1}(k)$ with the $q$th-order divergence at $k_s = \pi/2$, $\varphi_{q1}(k,t)$ corresponds to $q$th-order SS. By using the inverse Fourier transformation, we obtain the function of the $q$th-order lasing in the real space

$$\varphi_{q1}(j,t) = (2\pi)^{-1/2} \int_{-\pi}^{\pi} \varphi_{q1}(k,t) e^{ikj} dk, \quad (13)$$

that is,

$$\varphi_{q1}(j,t) = \frac{q! i^q e^{i\pi j/2}}{\frac{d^q s_{q1}^{-1}(k)}{dk^q}\Big|_{k=\pi/2} \sqrt[4]{\pi\sigma^2}} \int \cdots \int e^{-(j-N_t)^2/2\sigma^2} \times [d(j-N_t)]^q. \quad (14)$$

Appendix presents the details for the derivation of $\varphi_{q1}(j,t)$. It is shown that the profile of the $q$th-order lasing is the $q$th times integral of the Gaussian function. For different lasing orders, they have significantly different dynamic profiles; thus, the time evolution of the Gaussian wave packet distinguishes the order of SS.

To demonstrate our findings, we perform numerical simulations with a Gaussian wave packet injection. The initial state of a Gaussian wave packet on port 1 has the form of $(1/\sqrt[4]{\pi\sigma^2})\sum_j e^{-(j-N_c)^2/(2\sigma^2)} e^{-i\pi j/2} |j\rangle$, where $|j\rangle$ stands for the resonator mode $j$ in port 1. Figure 3 shows the profile of time evolution $|\varphi_{q1}(j,t)|$ in the outputs of the ports $q \in [1, 5]$, which exhibits the different dynamic features of high-order lasing. As expected, high-order lasing has enhanced wave emission. The analytical expressions of $\varphi_{q1}(j,t)$ in Appendix well describe the numerical simulations. From the analytical results, we emphasize the scaling law associated with high-order lasing in the form of

$$|\varphi_{q1}(j,t)| \begin{cases} \propto |j-N_t|^{q-1}, & (j < N_t) \\ \approx 0, & (j > N_t) \end{cases}. \quad (15)$$

Furthermore, we find that the total intensity $P_{q1}(t)$ in port $q$ satisfies

$$P_{q1}(t) = \sum_{j=0}^{\infty} |\varphi_{q1}(j,t)|^2 \propto t^{2q-1}. \quad (16)$$

Figure 2(c) shows the intensity growth at different high-order SSs. The intensity is polynomially enhanced through the scattering center at the EP; the intensity in the port at the first-order SS linearly increases with time $t$ [77], and the intensity in the scattering center at the $(q-p+1)$th-order EP is proportional to $t^{2(q-p)}$ [78]. Therefore, the lasing at the $(q-p+1)$th-order SS in port $q \geqslant p$ has the intensity $t^{2(q-p)+1}$ and $P(t) = 0$ in port $q < p$. The order of the growth in the polynomial intensity linearly scales with the order of the SS. This is the scaling law of the dynamics at the high-order SS. The different order of lasing dynamics exhibited from the high-order SS is attributed to the EP causing unidirectionality at the coalescence of SSs, which differs from the interference causing unidirectionality at the coincidence of SSs [29,30].

## V. THE EFFECT OF COHERENT INPUT

Furthermore, the order of lasing is controlled by the coherent input if the input vector equals the eigenstate or the generalized eigenstate because the generalized eigenstate of $h_c(k)$ or $H_c$ does not participate in the dynamics. From $E_N(k) = \Theta^{-1} h_c(k) \Theta$, we know that the columns of $\Theta$ have a set of $m$ linearly independent generalized eigenvectors associated with the eigenvalue $\epsilon_k$; $\Theta \chi_l$ is the generalized eigenstate of rank $(m+1-l)$ ($l = 1, 2, \ldots, m$), where $\chi_l = [\chi_l(1), \ldots, \chi_l(N)]^T$ with $\chi_l(j) = \delta_{lj}$. In this sense, the generalized eigenstate of rank 1 is the eigenstate of the system at the $m$th-order EP. Notably, $\Theta \chi_l$ is the generalized eigenstate of $h_c(k)$ and $H_c$ from $h_c(k) = H_c - (\omega_c - \kappa e^{-ik}) I_N$.

The dynamics of $h_c(k)$ are governed by its propagator

$$e^{-ih_c(k)t} = e^{-i\epsilon_k t} \sum_{j=0}^{+\infty} \frac{(-i)^j}{j!} [h_c(k) - \epsilon_k]^j t^j. \quad (17)$$

The input vector $\Theta \chi_{m+1-l}$ is the generalized eigenstate of rank $l$. Thus, we have $[h_c(k) - \epsilon_k]^l \Theta \chi_{m+1-l} = 0$ and $[h_c(k) - \epsilon_k]^{l-1} \Theta \chi_{m+1-l} \neq 0$. For the input vector $\Theta \chi_{m+1-l}$, the time evolution becomes

$$e^{-ih_c(k)t} \Theta \chi_{m+1-l} = e^{-i\epsilon_k t} \sum_{j=0}^{l-1} \frac{(-i)^j}{j!} [h_c(k) - \epsilon_k]^j t^j \propto t^{l-1}. \quad (18)$$

This describes the dynamics for the $l$th-order EP and the lasing enhancement caused by the $m$th-order EP is suppressed.

For the input vector $\Theta\chi_{m+1-l}$, the output vector $S(k)\Theta\chi_{m+1-l}$ is obtained in the form of

$$S(k)\Theta\chi_{m+1-l} = \left[-I_N - \frac{2i\kappa \sin k}{H_c - (\omega_c - \kappa e^{-ik})I_N}\right]\Theta\chi_{m+1-l}. \quad (19)$$

We simplify the output vector to obtain $S(k)\Theta\chi_{m+1-l} = -\Theta\chi_{m+1-l} - 2i\kappa \sin(k) h_c^{-1}(k)\Theta\chi_{m+1-l}$. Alternatively, the output vector $S(k)\Theta\chi_{m+1-l}$ for the input vector $\Theta\chi_{m+1-l}$ is also in the form of

$$S(k)\Theta\chi_{m+1-l} = -\Theta\chi_{m+1-l} - 2i\kappa \sin(k)\Theta E_N^{-1}(k)\chi_{m+1-l}. \quad (20)$$

The order of divergence in the output $S(k)\Theta\chi_{m+1-l}$ is determined by the dominated term $E_N^{-1}(k)\chi_{m+1-l}$ with the divergence, where the highest order of divergence is the $l$th order. Therefore, the input vector at the generalized eigenstate of rank $l$ changes the output into the $l$th-order divergence. The coherent input at the high-order SS alters the order of SS. This provides a unique way of lasing at the high-order SS. The different order of lasing dynamics exhibited from the high-order SS is attributed to the EP causing unidirectionality at the coalescence of SSs, which differs from the interference causing unidirectionality at the coincidence of SSs [29,30].

## VI. DISCUSSION

The proposed formation mechanism for the high-order SS as a guiding principle applies well in the coupled microcavity systems with fiber taper waveguides as the input and output channels [3,37]. In this case, we also discuss the analytical framework for the high-order SS. In the temporal coupled-mode theory, the scattering matrix is related to the Hamiltonian of the coupled microcavities $H_c$ by

$$S(\omega) = 1 - 2iD^\dagger \frac{1}{\omega - (H_c - iDD^\dagger)} D, \quad (21)$$

where $\omega$ is the frequency of the input and $D_{ij}$ is a matrix of coupling coefficients between microcavity mode $i$ and asymptotic waveguide channel $j$.

The term $(H_c - iDD^\dagger)$ defines the effective Hamiltonian $H_{\text{eff}}$ that involves the optical dissipation into the waveguide channels. Notably, $\omega - H_{\text{eff}}$ under a similar transformation $E_N(\omega) = \Theta'^{-1} h_c(k)\Theta'$ is either a diagonal canonical form for a diagonalizable $H_{\text{eff}}$ or a Jordan canonical form for a nondiagonalizable $H_{\text{eff}}$. The operator $\Theta'$ is $\omega$ independent because $H_{\text{eff}}$ is $\omega$ independent. The scattering coefficient divergence requires that the denominator of $S(\omega)$, i.e., $\omega - H_{\text{eff}}$, has an eigenvalue $\epsilon_\omega = 0$ at the critical $\omega = \omega_s$, and the $m$th-order divergence in $S(\omega)$ only appears when $H_{\text{eff}}$ is at the $m$th-order EP where $m$ eigenvalues of $\omega - H_{\text{eff}}$ coalesce to $\epsilon_\omega = 0$. We emphasize that the high-order SS in the coupled microcavity systems with fiber waveguides as the input and output channels can only be achieved when the effective Hamiltonian $H_{\text{eff}}$ is at the high-order EP.

We change $H_c$ at the high-order SS into its complex counterpart $H_c^*$ to obtain the high-order CPA. The second-order CPA is equivalent to the CPA EP [55,56], which can be well explained in our analytical framework. The effective Hamiltonian $H_{\text{eff}}$ at the second-order EP results in the second-order CPA and leads to a quartic absorption line shape. The higher order of CPA has a more flattened absorption line shape, which provides a highly enhanced absorbing capacity.

The time-reversed high-order laser is a high-order CPA. If the scattering matrix $S(k)$ acts on the incoming wave amplitudes $A = (a_1, a_2, \ldots, a_{N-1}, a_N)^T$, the outgoing wave amplitudes $B = (b_1, b_2, \ldots, b_{N-1}, b_N)^T$ are obtained as

$$B = S(k)A. \quad (22)$$

The $N$th-order laser appears at the $N$th-order SS, which is the $N$th-order pole of scattering matrix $S(k)$. Therefore, the $N$th-order divergence presents in the output vector $B$ for the zeroth-order input vector $A$ without the divergence. Consequently, the time-revered $N$th-order laser is a $N$th-order CPA. Notably, $[S^*(k)]^{-1}$ is the scattering matrix of the $N$th-order CPA

$$A^* = [S^*(k)]^{-1} B^*. \quad (23)$$

If the scattering matrix of the $N$th-order CPA acts on the incoming wave amplitudes $B^*$ with the $N$th-order divergence, the outgoing wave amplitudes $A^*$ without the divergence are obtained. The $N$th-order CPA can remove the divergence in the incident wave by $N$ orders. For coherent wave emissions from the lower-order lasers, the order of divergence is less than $N$. Therefore, the high-order CPA can completely absorb coherent wave emissions from the lower-order lasers. The high-order CPA perfectly absorbs coherent wave emissions from the lower-order lasers, indicating greatly enhanced absorption capacity. Notably, the high-order CPA cannot be implemented if the CPAs are connected in series; thus, the high-order CPA is irreplaceable in optical engineering.

## VII. CONCLUSION

The high-order SS stimulates new opportunities in photonics. We have demonstrated the formation of high-order SS and high-order lasing. The EP and SS are unified at the high-order SS, where the wave emission intensity accumulates in a manner described by a power law. Notably, the input superposed at the eigenstate or generalized eigenstate of the scattering center alters the order of wave emission due to the varying participation of the EP. In addition, the time-reversed high-order lasers are high-order CPAs that can perfectly absorb coherent wave emissions from lower-order lasers. The conclusions for the coupled resonators [79] also apply to coupled waveguides [80] and other platforms [81,82]. The realization of high-order lasers and CPAs in nonlinear systems is promising [58,83].

## ACKNOWLEDGMENTS

We acknowledge Huanan Li for useful discussions. This work was supported by the National Natural Science Foundation of China (Grant No. 12222504).

**APPENDIX: ANALYTICAL SOLUTION OF HIGH-ORDER LASING**

From Eq. (13), we have

$$\varphi_{q1}(j,t) = \frac{1}{\sqrt{2\pi}} \int_{-\pi}^{\pi} \varphi(k,t) s_{q1}(k) e^{ikj} dk, \tag{A1}$$

$$\approx \frac{1}{\sqrt{2\pi}} \int_{-\pi}^{\pi} \frac{\sqrt{\sigma}}{\sqrt[4]{\pi}} e^{-\frac{\sigma^2(k-\pi/2)^2}{2}} e^{-i(k-\pi/2)N_t} \frac{q!}{\frac{d^q s_{q1}^{-1}(k)}{dk^q}\Big|_{k=\pi/2}(k-\pi/2)^q} e^{ikj} dk, \tag{A2}$$

$$= \frac{q! i^q}{\frac{d^q s_{q1}^{-1}(k)}{dk^q}\Big|_{k=\pi/2}} \frac{\sqrt{\sigma}}{\sqrt[4]{4\pi^3}} e^{i\pi j/2} \int_{-\pi}^{\pi} e^{-\frac{\sigma^2(k-\pi/2)^2}{2}} e^{i(k-\pi/2)(j-N_t)} \frac{1}{i^q(k-\pi/2)^q} dk. \tag{A3}$$

We set the integral

$$I(j - N_t) = \int_{-\pi}^{\pi} e^{-\frac{\sigma^2(k-\pi/2)^2}{2}} e^{i(k-\pi/2)(j-N_t)} \frac{1}{i^q(k-\pi/2)^q} dk. \tag{A4}$$

Then, we find

$$\frac{d^q I(j - N_t)}{[d(j-N_t)]^q} = \int_{-\pi}^{\pi} e^{-\frac{\sigma^2(k-\pi/2)^2}{2}} e^{i(k-\pi/2)(j-N_t)} dk = \sqrt{\frac{2\pi}{\sigma^2}} e^{-\frac{(j-N_t)^2}{2\sigma^2}}, \tag{A5}$$

and we obtain the final expression in Eq. (14).

Applying the condition $\varphi_{q1}(\infty, t) = 0$, we can obtain the analytical expression of $\varphi_{q1}(j,t)$ with $q \in [1, 5]$. The expressions from $\varphi_{11}(j,t)$ to $\varphi_{51}(j,t)$ correspond to the lasing from the first order to the fifth order.

$$\varphi_{11}(j,t) = \xi \,\text{erfc}\left(\frac{j - N_t}{\sqrt{2}\sigma}\right) e^{i\pi j/2}, \tag{A6}$$

$$\varphi_{21}(j,t) = -i\xi \left[(j - N_t)\text{erfc}\left(\frac{j-N_t}{\sqrt{2}\sigma}\right) - \sqrt{\frac{2}{\pi}} \sigma e^{-\frac{(j-N_t)^2}{2\sigma^2}}\right] e^{i\pi j/2}, \tag{A7}$$

$$\varphi_{31}(j,t) = -\xi \left[\frac{(j-N_t)^2 + \sigma^2}{2}\text{erfc}\left(\frac{j-N_t}{\sqrt{2}\sigma}\right) - \frac{\sigma}{\sqrt{2\pi}}(j-N_t) e^{-\frac{(j-N_t)^2}{2\sigma^2}}\right] e^{i\pi j/2}, \tag{A8}$$

$$\varphi_{41}(j,t) = i\xi \left[\frac{(j-N_t)^3 + 3\sigma^2(j-N_t)}{6}\text{erfc}\left(\frac{j-N_t}{\sqrt{2}\sigma}\right) - \sigma \frac{(j-N_t)^2 + 2\sigma^2}{3\sqrt{2\pi}} e^{-\frac{(j-N_t)^2}{2\sigma^2}}\right] e^{i\pi j/2}, \tag{A9}$$

$$\varphi_{51}(j,t) = \xi \left[\frac{(j-N_t)^4 + 6\sigma^2(j-N_t)^2 + 3\sigma^4}{24}\text{erfc}\left(\frac{j-N_t}{\sqrt{2}\sigma}\right) - \sqrt{2}\sigma \frac{(j-N_t)^3 + 5\sigma^2(j-N_t)}{\sqrt{\pi}} e^{-\frac{(j-N_t)^2}{2\sigma^2}}\right] e^{i\pi j/2}, \tag{A10}$$

where $\xi = \sqrt[4]{4\pi\sigma^2}$ and erfc is the complementary error function.

[1] W. D. Heiss, Exceptional points of non-Hermitian operators, J. Phys. A: Math. Gen. **37**, 2455 (2004).

[2] C. E. Rüter, K. G. Makris, R. El-Ganainy, D. N. Christodoulides, M. Segev, and D. Kip, Observation of parity-time symmetry in optics, Nat. Phys. **6**, 192 (2010).

[3] B. Peng, S. K. Özdemir, F. Lei, F. Monifi, M. Gianfreda, G. L. Long, S. Fan, F. Nori, C. M. Bender, and L. Yang, Parity-time-symmetric whispering-gallery microcavities, Nat. Phys. **10**, 394 (2014).

[4] T. Goldzak, A. A. Mailybaev, and N. Moiseyev, Light Stops at Exceptional Points, Phys. Rev. Lett. **120**, 013901 (2018).

[5] L. J. Fernández-Alcázar, H. Li, F. Ellis, A. Alù, and T. Kottos, Robust Scattered Fields from Adiabatically Driven Targets around Exceptional Points, Phys. Rev. Lett. **124**, 133905 (2020).

[6] A. Mostafazadeh, Spectral Singularities of Complex Scattering Potentials and Infinite Reflection and Transmission Coefficients at Real Energies, Phys. Rev. Lett. **102**, 220402 (2009).

[7] A. Mostafazadeh, Nonlinear Spectral Singularities for Confined Nonlinearities, Phys. Rev. Lett. **110**, 260402 (2013).

[8] S. Longhi, Spectral singularities in a non-Hermitian Friedrichs-Fano-Anderson model, Phys. Rev. B **80**, 165125 (2009).

[9] S. Longhi, $\mathcal{PT}$-symmetric laser absorber, Phys. Rev. A **82**, 031801(R) (2010).

[10] Y. D. Chong, L. Ge, and A. D. Stone, $\mathcal{PT}$-Symmetry Breaking and Laser-Absorber Modes in Optical Scattering Systems, Phys. Rev. Lett. **106**, 093902 (2011).

[11] L. Ge, Y. D. Chong, and A. D. Stone, Conservation relations and anisotropic transmission resonances in one-dimensional $\mathcal{PT}$-symmetric photonic heterostructures, Phys. Rev. A **85**, 023802 (2012).

[12] X. Liu, S. D. Gupta, and G. S. Agarwal, Regularization of the spectral singularity in $\mathcal{PT}$-symmetric systems by all-order

nonlinearities: Nonreciprocity and optical isolation, Phys. Rev. A **89**, 013824 (2014).

[13] C. Hang, G. Huang, and V. V. Konotop, Tunable spectral singularities: Coherent perfect absorber and laser in an atomic medium, New J. Phys. **18**, 085003 (2016).

[14] S. Pendharker, Y. Guo, F. Khosravi, and Z. Jacob, $\mathcal{PT}$-symmetric spectral singularity and negative-frequency resonance, Phys. Rev. A **95**, 033817 (2017).

[15] V. V. Konotop and D. A. Zezyulin, Spectral singularities of odd-PT-symmetric potentials, Phys. Rev. A **99**, 013823 (2019).

[16] V. V. Konotop, E. Lakshtanov, and B. Vainberg, Designing lasing and perfectly absorbing potentials, Phys. Rev. A **99**, 043838 (2019).

[17] Y. Li and C. Argyropoulos, Exceptional points and spectral singularities in active epsilon-near-zero plasmonic waveguides, Phys. Rev. B **99**, 075413 (2019).

[18] Q. Liu, C. Qin, B. Wang, and P. Lu, Scattering singularities of optical waveguides under complex modulation, Phys. Rev. A **101**, 033818 (2020).

[19] H. Ramezani, Spectral singularities with directional sensitivity, Phys. Rev. A **103**, 043516 (2021).

[20] Z. Sakotic, A. Krasnok, A. Alù, and N. Jankovic, Topological scattering singularities and embedded eigenstates for polarization control and sensing applications, Photon. Res. **9**, 1310 (2021).

[21] H. Schomerus, Fundamental constraints on the observability of non-Hermitian effects in passive systems, Phys. Rev. A **106**, 063509 (2022).

[22] M. A. Simón, A. Buendía, A. Kiely, A. Mostafazadeh, and J. G. Muga, S-matrix pole symmetries for non-Hermitian scattering Hamiltonians, Phys. Rev. A **99**, 052110 (2019).

[23] Y. D. Chong, L. Ge, H. Cao, and A. D. Stone, Coherent Perfect Absorbers: Time-Reversed Lasers, Phys. Rev. Lett. **105**, 053901 (2010).

[24] W. Wan, Y. Chong, L. Ge, H. Noh, A. D. Stone, and H. Cao, Time-reversed lasing and interferometric control of absorption, Science **331**, 889 (2011).

[25] J. R. Piper, V. Liu, and S. Fan, Total absorption by degenerate critical coupling, Appl. Phys. Lett. **104**, 251110 (2014).

[26] D. G. Baranov, A. Krasnok, T. Shegai, A. Alù, and Y. Chong, Coherent perfect absorbers: Linear control of light with light, Nat. Rev. Mater. **2**, 17064 (2017).

[27] H. Li, S. Suwunnarat, R. Fleischmann, H. Schanz, and T. Kottos, Random Matrix Theory Approach to Chaotic Coherent Perfect Absorbers, Phys. Rev. Lett. **118**, 044101 (2017).

[28] J. Jeffers, Nonlocal Coherent Perfect Absorption, Phys. Rev. Lett. **123**, 143602 (2019).

[29] H. Ramezani, H.-K. Li, Y. Wang, and X. Zhang, Unidirectional Spectral Singularities, Phys. Rev. Lett. **113**, 263905 (2014).

[30] L. Jin and Z. Song, Incident Direction Independent Wave Propagation and Unidirectional Lasing, Phys. Rev. Lett. **121**, 073901 (2018).

[31] L. Feng, R. El-Ganainy, and L. Ge, Non-Hermitian photonics based on parity-time symmetry, Nat. Photonics **11**, 752 (2017).

[32] R. El-Ganainy, K. G. Makris, M. Khajavikhan, Z. H. Musslimani, S. Rotter, and D. N. Christodoulides, Non-Hermitian physics and PT symmetry, Nat. Phys. **14**, 11 (2018).

[33] M.-A. Miri and A. Alù, Exceptional points in optics and photonics, Science **363**, eaar7709 (2019).

[34] S. K. Özdemir, S. Rotter, F. Nori, and L. Yang, Parity-time symmetry and exceptional points in photonics, Nat. Mater. **18**, 783 (2019).

[35] Q. Liu, S. Li, B. Wang, S. Ke, C. Qin, K. Wang, W. Liu, D. Gao, P. Berini, and P. Lu, Efficient Mode Transfer on a Compact Silicon Chip by Encircling Moving Exceptional Points, Phys. Rev. Lett. **124**, 153903 (2020).

[36] K. Ding, C. Fang, and G. Ma, Non-Hermitian topology and exceptional-point geometries, Nat. Rev. Phys. **4**, 745 (2022).

[37] J. Wiersig, Enhancing the Sensitivity of Frequency and Energy Splitting Detection by Using Exceptional Points: Application to Microcavity Sensors for Single-Particle Detection, Phys. Rev. Lett. **112**, 203901 (2014).

[38] Z.-P. Liu, J. Zhang, S. K. Özdemir, B. Peng, H. Jing, X.-Y. Lü, C.-W. Li, L. Yang, F. Nori, and Y.-X. Liu, Metrology with $\mathcal{PT}$-symmetric cavities: Enhanced sensitivity near the PT-phase transition, Phys. Rev. Lett. **117**, 110802 (2016).

[39] W. Chen, S. K. Özdemir, G. Zhao, J. Wiersig, and L. Yang, Exceptional points enhance sensing in an optical microcavity, Nature (London) **548**, 192 (2017).

[40] H. Hodaei, A. U. Hassan, S. Wittek, H. Garcia-Gracia, R. El-Ganainy, D. N. Christodoulides, and M. Khajavikhan, Enhanced sensitivity at higher-order exceptional points, Nature (London) **548**, 187 (2017).

[41] H.-K. Lau and A. A. Clerk, Fundamental limits and nonreciprocal approaches in non-Hermitian quantum sensing, Nat. Commun. **9**, 4320 (2018).

[42] N. A. Mortensen, P. A. D. Gonçalves, M. Khajavikhan, D. N. Christodoulides, C. Tserkezis, and C. Wolff, Fluctuations and noise-limited sensing near the exceptional point of parity-time-symmetric resonator systems, Optica **5**, 1342 (2018).

[43] Z. Xiao, H. Li, T. Kottos, and A. Alù, Enhanced Sensing and Nondegraded Thermal Noise Performance Based on $\mathcal{PT}$-Symmetric Electronic Circuits with a Sixth-Order Exceptional Point, Phys. Rev. Lett. **123**, 213901 (2019).

[44] M. Zhang, W. Sweeney, C. W. Hsu, L. Yang, A. D. Stone, and L. Jiang, Quantum Noise Theory of Exceptional Point Amplifying Sensors, Phys. Rev. Lett. **123**, 180501 (2019).

[45] Q. Zhong, J. Ren, M. Khajavikhan, D. N. Christodoulides, S. K. Özdemir, and R. El-Ganainy, Sensing with Exceptional Surfaces in Order to Combine Sensitivity with Robustness, Phys. Rev. Lett. **122**, 153902 (2019).

[46] Y.-H. Lai, Y.-K. Lu, M.-G. Suh, Z. Yuan, and K. Vahala, Observation of the exceptional-point-enhanced Sagnac effect, Nature (London) **576**, 65 (2019).

[47] M. P. Hokmabadi, A. Schumer, D. N. Christodoulides, and M. Khajavikhan, Non-Hermitian ring laser gyroscopes with enhanced Sagnac sensitivity, Nature (London) **576**, 70 (2019).

[48] A. McDonald and A. A. Clerk, Exponentially-enhanced quantum sensing with non-Hermitian lattice dynamics, Nat. Commun. **11**, 5382 (2020).

[49] J. Doppler, A. A. Mailybaev, J. Böhm, U. Kuhl, A. Girschik, F. Libisch, T. J. Milburn, P. Rabl, N. Moiseyev, and S. Rotter, Dynamically encircling an exceptional point for asymmetric mode switching, Nature (London) **537**, 76 (2016).

[50] H. Xu, D. Mason, L. Jiang, and J. G. E. Harris, Topological energy transfer in an optomechanical system with exceptional points, Nature (London) **537**, 80 (2016).

[51] B. Peng, S. K. Özdemir, M. Liertzer, W. Chen, J. Kramer, H. Yılmaz, J. Wiersig, S. Rotter, and L. Yang, Chiral modes and directional lasing at exceptional points, Proc. Natl. Acad. Sci. USA **113**, 6845 (2016).

[52] Z. Lin, H. Ramezani, T. Eichelkraut, T. Kottos, H. Cao, and D. N. Christodoulides, Unidirectional Invisibility Induced by $\mathcal{PT}$-Symmetric Periodic Structures, Phys. Rev. Lett. **106**, 213901 (2011).

[53] L. Feng, Y.-L. Xu, W. S. Fegadolli, M.-H. Lu, J. E. B. Oliveira, V. R. Almeida, Y.-F. Chen, and A. Scherer, Experimental demonstration of a unidirectional reflectionless parity-time metamaterial at optical frequencies, Nat. Mater. **12**, 108 (2013).

[54] R. Fleury, D. Sounas, and A. Alù, An invisible acoustic sensor based on parity-time symmetry, Nat. Commun. **6**, 5905 (2015).

[55] W. R. Sweeney, C. W. Hsu, S. Rotter, and A. D. Stone, Perfectly Absorbing Exceptional Points and Chiral Absorbers, Phys. Rev. Lett. **122**, 093901 (2019).

[56] C. Wang, W. R. Sweeney, A. D. Stone, and L. Yang, Coherent perfect absorption at an exceptional point, Science **373**, 1261 (2021).

[57] S. Soleymani, Q. Zhong, M. Mokim, S. Rotter, R. El-Ganainy, and S. K. Özdemir, Chiral and degenerate perfect absorption on exceptional surfaces, Nat. Commun. **13**, 599 (2022).

[58] S. Suwunnarat, Y. Tang, M. Reisner, F. Mortessagne, U. Kuhl, and T. Kottos, Non-linear coherent perfect absorption in the proximity of exceptional points, Commun. Phys. **5**, 5 (2022).

[59] Z. Sakotic, P. Stankovic, V. Bengin, A. Krasnok, A. Alù, and N. Jankovic, Non-Hermitian control of topological scattering singularities emerging from bound states in the continuum, Laser Photonics Rev. **10**, 2200308 (2023).

[60] A. Yariv, Coupled-mode theory for guided-wave optics, IEEE J. Quantum Electron. **9**, 919 (1973).

[61] D. Marcuse, Coupled mode theory of optical resonant cavities, IEEE J. Quantum Electron. **21**, 1819 (1985).

[62] J. D. Joannopoulos, S. G. Johnson, J. N. Winn, and R. D. Meade, *Photonic Crystals: Molding the Flow of Light* (Princeton University Press, Princeton, 2008).

[63] A. Krasnok, D. Baranov, H. Li, M.-A. Miri, F. Monticone, and A. Alù, Anomalies in light scattering, Adv. Opt. Photon. **11**, 892 (2019).

[64] W. R. Sweeney, C. W. Hsu, and A. D. Stone, Theory of reflectionless scattering modes, Phys. Rev. A **102**, 063511 (2020).

[65] L. Jin and Z. Song, Partitioning technique for discrete quantum systems, Phys. Rev. A **83**, 062118 (2011).

[66] V. V. Konotop, B. C. Sanders, and D. A. Zezyulin, Spectral singularities of a potential created by two coupled microring resonators, Opt. Lett. **44**, 2024 (2019).

[67] D. A. Zezyulin and V. V. Konotop, Universal form of arrays with spectral singularities, Opt. Lett. **45**, 3447 (2020).

[68] K. Ding, G. Ma, M. Xiao, Z. Q. Zhang, and C. T. Chan, Emergence, Coalescence, and Topological Properties of Multiple Exceptional Points and Their Experimental Realization, Phys. Rev. X **6**, 021007 (2016).

[69] M. P. Hokmabadi, N. S. Nye, R. El-Ganainy, D. N. Christodoulides, and M. Khajavikhan, Supersymmetric laser arrays, Science **363**, 623 (2019).

[70] B. Midya, H. Zhao, X. Qiao, P. Miao, W. Walasik, Z. Zhang, N. M. Litchinitser, and L. Feng, Supersymmetric microring laser arrays, Photonics Res. **7**, 363 (2019).

[71] Q. Zhong, J. Kou, S. K. Özdemir, and R. El-Ganainy, Hierarchical Construction of Higher-Order Exceptional Points, Phys. Rev. Lett. **125**, 203602 (2020).

[72] Y.-X. Xiao, Z.-Q. Zhang, Z. H. Hang, and C. T. Chan, Anisotropic exceptional points of arbitrary order, Phys. Rev. B **99**, 241403(R) (2019).

[73] J. Ren, Y. G. N. Liu, M. Parto, W. E. Hayenga, M. P. Hokmabadi, D. N. Christodoulides, and M. Khajavikhan, Unidirectional light emission in $\mathcal{PT}$-symmetric microring lasers, Opt. Express **26**, 27153 (2018).

[74] Q. Zhong, S. K. Özdemir, A. Eisfeld, A. Metelmann, and R. El-Ganainy, Exceptional-Point-Based Optical Amplifiers, Phys. Rev. Appl. **13**, 014070 (2020).

[75] A. Muñoz de las Heras, R. Franchi, S. Biasi, M. Ghulinyan, L. Pavesi, and I. Carusotto, Nonlinearity-Induced Reciprocity Breaking in a Single Nonmagnetic Taiji Resonator, Phys. Rev. Appl. **15**, 054044 (2021).

[76] C. Yuce and H. Ramezani, Robust exceptional points in disordered systems, Europhys. Lett. **126**, 17002 (2019).

[77] P. Wang, L. Jin, G. Zhang, and Z. Song, Wave emission and absorption at spectral singularities, Phys. Rev. A **94**, 053834 (2016).

[78] Q. Zhong, D. N. Christodoulides, M. Khajavikhan, K. G. Makris, and R. El-Ganainy, Power-law scaling of extreme dynamics near higher-order exceptional points, Phys. Rev. A **97**, 020105(R) (2018).

[79] M. Parto, S. Wittek, H. Hodaei, G. Harari, M. A. Bandres, J. Ren, M. C. Rechtsman, M. Segev, D. N. Christodoulides, and M. Khajavikhan, Edge-Mode Lasing in 1D Topological Active Arrays, Phys. Rev. Lett. **120**, 113901 (2018).

[80] M. Pan, H. Zhao, P. Miao, S. Longhi, and L. Feng, Photonic zero mode in a non-Hermitian photonic lattice, Nat. Commun. **9**, 1308 (2018).

[81] W. Gou, T. Chen, D. Xie, T. Xiao, T.-S. Deng, B. Gadway, W. Yi, and B. Yan, Tunable Nonreciprocal Quantum Transport through a Dissipative Aharonov-Bohm Ring in Ultracold Atoms, Phys. Rev. Lett. **124**, 070402 (2020).

[82] L. Ding, K. Shi, Q. Zhang, D. Shen, X. Zhang, and W. Zhang, Experimental Determination of $\mathcal{PT}$-Symmetric Exceptional Points in a Single Trapped Ion, Phys. Rev. Lett. **126**, 083604 (2021).

[83] A. Müllers, B. Santra, C. Baals, J. Jiang, J. Benary, R. Labouvie, D. A. Zezyulin, V. V. Konotop, and H. Ott, Coherent perfect absorption of nonlinear matter waves, Sci. Adv. **4**, eaat6539 (2018).